\documentclass[12pt]{article}
\usepackage{amssymb}
\usepackage{graphicx}
\usepackage{latexsym}
\newcommand{\eq}{\begin{equation}}
\newcommand{\en}{\end{equation}}
\newcommand{\qe}{\end{equation}}
\newcommand{\ear}{\begin{eqnarray}}
\newcommand{\eqa}{\begin{eqnarray}}
\newcommand{\rae}{\end{eqnarray}}
\newcommand{\ena}{\end{eqnarray}}
\newcommand{\beq}{\begin{equation}} 
\newcommand{\eeq}{\end{equation}}
\newcommand{\bea}{\begin{eqnarray}}
\newcommand{\eea}{\end{eqnarray}}
\newcommand{\Z}{{\sf Z \!\!\! Z}}
\newcommand{\twolink}{{\mathbb{T}}}
\begin{document}
\begin{titlepage}
\vskip0.5cm
\begin{flushright}
DFTT 13/2004\\
IFUM-795-FT
\end{flushright}
\vskip0.5cm
\begin{center} 
{\Large\bf  Static quark potential and effective string corrections in
  the (2+1)-d SU(2) Yang-Mills theory }
\end{center}
\vskip1.3cm
\centerline{
M. Caselle$^a$, M. Pepe$^{b}$ and A. Rago$^c$}
 \vskip1.0cm
 \centerline{\sl  $^a$ Dipartimento di Fisica
 Teorica dell'Universit\`a di Torino and I.N.F.N.,}
 \centerline{\sl via P.Giuria 1, I-10125 Torino, Italy}
 \centerline{\sl e--mail:  caselle@to.infn.it }
 \vskip0.4 cm
 \centerline{\sl  $^b$ Institute for Theoretical Physics, 
Bern University,} 
 \centerline{\sl Sidlerstrasse 5, CH-3012 Bern, Switzerland}
 \centerline{\sl e--mail: pepe@itp.unibe.ch}
 \vskip0.4 cm
 \centerline{\sl  $^c$ Dipartimento di Fisica
dell'Universit\`a di Milano and I.N.F.N.,}
 \centerline{\sl via Celoria 16, I-20133 Milano, Italy}
 \centerline{\sl e--mail: antonio.rago@mi.infn.it}
 \vskip1.0cm

\begin{abstract}
We report on a very accurate measurement of the static quark potential in
$SU(2)$ Yang-Mills theory in (2+1) dimensions in order to study the corrections to the
linear behaviour. We perform numerical simulations at zero and finite
temperature comparing our results with the corrections given by the effective string
picture in these two regimes. We also check for universal features discussing our results
together with those recently published for the (2+1)-d $\Z(2)$ and $SU(3)$ pure gauge
theories.     
\end{abstract}
\end{titlepage}

\setcounter{footnote}{0}
\def\thefootnote{\arabic{footnote}}

One of the most interesting open problems in lattice gauge theories is the construction of
an effective string description of the static quark potential. Starting from the seminal
papers of L\"uscher, Symanzik and Weisz~\cite{lsw,Lus81} much progress has been done in this
direction. On the analytical side, the corrections to the linear term of the static quark
potential induced by an effective string description have been calculated. These
computations have been carried out considering effective string actions with the most
general quartic self-interaction term and any type of boundary conditions for a
rectangular $L\times R$ string world-sheet~\cite{df83}. Recently a computation of the
subleading correction has been performed also for the case of the three static
quarks~\cite{Jah03}. On the numerical side, the main 
prediction of the effective string theory, i.e. the presence in the static quark potential
of the so called ``L\"uscher term", has been tested in several pure gauge
theories, ranging from the 3-d $\Z(2)$ gauge model to the 4-d $SU(3)$ Yang-Mills
theory~\cite{cfghp}-\cite{maj03}. String effects have also been investigated in the excited
states of the static quark potential: these studies indicate that an effective string
description seems to hold when the distance between the quark and the antiquark is about
1.5-2$fm$ \cite{jkm02,jkm03,Jug04,jkm03b}. It is still to be understood how these results match with
those of the ground state where the effective string description turns out to be valid
already at about 0.5$fm$.

In the last two years the development of new numerical methods -- mainly the L\"uscher
and Weisz's multilevel algorithm~\cite{lw01} -- has allowed to test with much higher
accuracy the effective string picture of the static quark potential. Furthermore, with the
increasing precision of numerical results, new important issues concerning higher order
terms in the effective string action can be addressed. 

Numerical studies have shown that at short distances and at high temperatures (but
still in the confined phase) the naive picture of a free bosonic effective string is
inadequate, however it is not clear how it should be modified. The simplest possible
proposal, i.e a Nambu-Goto like action, seems to describe rather well the large
distance/high temperature behaviour of the static quark potential~\cite{cpp02,chp03,chp04,chp04b}, but
it seems to be less successful in the short distance/low temperature
regime~\cite{jkm03,chp04c,chp04}. Moreover, contrary to the L\"uscher term -- which is universal
and does not depend on the gauge group but only on the space-time dimensionality and on
the string boundary conditions -- there is no evidence (and no reason) that a similar
universality holds also for the higher order corrections.  It is thus interesting to
understand to what extent these higher order terms depend on the gauge group of the
theory.

In this paper we investigate these questions by studying the $SU(2)$ Yang-Mills theory in
$(2+1)$ dimensions. We have extracted the static quark potential from the 2-point
correlation function of Polyakov loops both at very low $(T \leq T_c/7)$ and at high
temperatures $(T=3T_c/4)$. We have then compared our results in these two regimes with the
predictions of the free bosonic string model and of the Nambu-Goto effective string model
truncated at the second perturbative order. Moreover we have also compared our results
with those obtained in~\cite{lw02} for the $SU(3)$ Yang-Mills theory and
in~\cite{chp03,chp04} for the $\Z (2)$ gauge theory in (2+1) dimensions. In this study we
have addressed the following three issues: 
\begin{itemize}
\item quantify the deviations from  the L\"uscher term in the short distance/low temperature
  regime and test if they are compatible with a Nambu-Goto type action and/or to the
  presence of a boundary term.  
\item study the large distance/high temperature regime of the static quark potential to
  check the reliability of the free bosonic string approximation as well as of the
  Nambu-Goto action. Notice that this is a non trivial test of the effective string
  description. In fact, in this regime, the effective string predictions can be obtained
  by a modular transformation of the short distance/low temperature results.
\item compare our results with the corrections measured in other pure gauge theories to
  study the presence of possible common features.   
\end{itemize}

This paper is organized as follows. In section~\ref{sect1} we discuss the main features of
the effective string description of the static quark potential, listing a set of
predictions which we shall later compare with the results of our numerical simulations. 
Section~\ref{sect2} is devoted to a general discussion of the (2+1)-d $SU(2)$ Yang-Mills
theory while in section~\ref{sect3} we give some details on the L\"uscher and Weisz's
algorithm we have used to generate our data. Then, in the second part of the paper, we 
compare the theoretical predictions with our numerical data: first (section~\ref{sect4})
at short distance/low temperature and then (section~\ref{sect5}) in the large
distance/high temperature regime. Finally, in section~\ref{sect6}, we compare the $\Z(2)$,
$SU(2)$ and $SU(3)$ results in the short distance/low temperature regime. Our
conclusions are contained in section~\ref{sect7}. A preliminary account of the results
presented here can be found in~\cite{cpp_proc}.

\section{Effective string predictions}
\label{sect1}
In this section we summarize some known results on the effective string description of the
static quark potential. For a more detailed discussion we refer the interested reader to
the original papers~\cite{lsw,Lus81,df83} or the recent review~\cite{chp03}. 

In a pure gauge theory the interaction between two static charges at temperature $T=1/L$
($L$ being the lattice size in the periodic temporal direction) can be extracted by
measuring the correlation function $G(R)$ of two Polyakov loops $P$ at distance $R$ 

\eq
G(R)\equiv 
\langle P(x)P (x+R)^* \rangle \equiv {\rm e}^{-F(R,L)}~~~.
\label{polya}
\en
The free energy $F(R,L)$ of the static quark-antiquark pair is then related to the static
quark potential $V(R)$ by 
\eq\label{statpot}
V(R) \equiv \frac{1}{L} F(R,L) = - \frac{1}{L} \log G(R)
\en
In the confined phase the force lines of the static quark-antiquark interaction are
focused into a flux tube connecting the two charges. For not too short distances, the
static potential rises linearly with a slope $\sigma$ given by the string tension: the
leading behavior of $F(R,L)$ is the so-called ``area law''
\eq
F(R,L)\sim \sigma L R + k(L)
~~~,\label{area}
\en
where $k(L)$ is a constant term depending only on $L$ and with no physical meaning. In the
strong coupling phase of the model -- i.e. below the roughening transition -- this
behaviour can be obtained analytically by a strong coupling expansion. In the rough phase
the transverse degrees of freedom of the flux tube become massless: the flux tube
delocalizes and the above description represents only the leading behaviour. The
suggestion of~\cite{lsw} is to try  an
effective description of the dynamics of the flux tube in terms of a fluctuating
string. According to this effective string picture, in the rough phase of the theory one
should modify eq.(\ref{area}) in order to take into account the quantum fluctuations of
the flux tube (``effective string corrections''). The exact form of these subleading
corrections is unknown: they are expected to be a complicated function of $R$ and $L$ and
they depend on the effective string action describing the flux tube dynamics. However,
assuming that the string is smooth and the self-interactions are very weak and negligible in first
approximation, one can consider the free bosonic string limit. This limit corresponds to
take into account only the leading term of the perturbative derivative expansion 
of the effective string action, dropping out all the other terms depending on the flux
tube dynamics. Hence the free bosonic string approximation amounts to consider a purely
geometrical description of the flux tube. Indeed one finds that the corrections
$F_1^q(R,L)$ to the area law behavior  
\eq
F(R,L)\sim  \sigma L R + k(L)+ F_1^q(R,L)
~~~\label{a+q}
\en
only depend on the shape and on the boundary conditions of the string world-sheet and on the
number, $(d-2)$, of the transverse dimensions. In the particular case in which we are
interested in this paper  -- i.e. periodic boundary conditions in the compactified temporal
direction and fixed spatial boundary conditions along the two Polyakov loops -- one finds: 
\eq
F_1^q(R,L)=(d-2)\log\left({\eta(\tau)}\right)
\hskip0.5cm
;\hskip0.5cm {-i}\tau={L\over 2R}~~~,
\label{bos}
\en
\noindent
where $\eta$ denotes the Dedekind eta function
\eq\label{etafun}
\eta(\tau)=q^{1\over24}\prod_{n=1}^\infty(1-q^n)\hskip0.5cm
;\hskip0.5cmq=e^{2\pi i\tau}~~~,
\en
The labels $q$ and $1$ in $F_1^q$ recall this to be the first order term in the
perturbative derivative expansion of the quantum fluctuations around the free string
approximation. Eq. (\ref{bos}) is commonly referred to as the ``free bosonic string
approximation''. 

The Dedekind function has different expansions in the two regimes $2R<L$ and  $2R>L$,
related to each other by the modular transformation 
$\eta(-1/\tau) = \sqrt{-i \tau}\,\eta(\tau)$. This yields to the following two different
expressions for the quantum corrections 
\begin{description}
\item{$\bullet$ $2R<L$}
\eq
F_1^q(R,L)=\left[-\frac{\pi L}{24 R}
+\sum_{n=1}^\infty \log (1-e^{-\pi nL/R})\right](d-2)~~~,
\label{zsmalltot}
\en
\item{$\bullet$ $2R>L$}
\eq
F_1^q(R,L)=\left[-\frac{\pi R}{6 L}+\frac{1}{2} \log\frac{2R}{L}
+\sum_{n=1}^\infty \log (1-e^{-4\pi nR/L})\right](d-2)~~~.
\label{zbigtot}
\en
\end{description}
The exponentially small corrections coming from the subleading terms in
eq.~(\ref{zsmalltot}) are negligible unless we are in the intermediate region $R\sim L/2$.
In eq. (\ref{zsmalltot}) the leading term of the expansion is the well known ``L\"uscher
term''~\cite{lsw} which decreases with the distance as $1/R$. Instead, in eq.
(\ref{zbigtot}), the first term is proportional to $R$ and to $T=1/L$: it is a finite
temperature correction which lowers the string tension as the temperature increases.
Interestingly, the expression obtained for $2R>L$, can also be reinterpreted as a low
temperature result. If the temporal direction is looked at as a spatial direction and
vice-versa, the correlator between the two Polyakov loops becomes the temporal evolution
for a time $R$ of a torelon of length $L$. The difference now is that the spatial
boundary conditions of the fluctuating string are periodic and no longer fixed. With these
different spatial boundary conditions the coefficient of the L\"uscher term turns out to
be 4 times larger~\cite{df83} as one can read out from the expansion (\ref{zbigtot}).

In our study we want to compare the expectations coming from effective string descriptions
of the static quark potential with the numerical results of Monte Carlo simulations. This
comparison will be performed considering the two different regimes that we have discussed
here above. In order to set the terminology, we refer to $2R<L$ as the {\em short
distance/low $T$} regime and to $2R>L$ as the {\em long distance/high $T$} regime. Note
that we are always in the confined phase of the pure gauge theory: hence, here, high $T$ 
means that we are at a finite temperature not far below the temperature of the
deconfinement phase transition.

At high enough temperatures (i.e. for small values of $L$), higher order terms of the
perturbative derivative expansion become important and they are no longer
negligible. These terms encode the string self-interaction and depend on the particular
choice of the effective string action. The simplest proposal -- discussed for instance 
in~\cite{df83,cpp02,chp03} -- is the Nambu--Goto action in which the string configurations
are weighted proportionally to the world-sheet area. In this model the next-to-leading
contribution to the free energy turns out to be (we have set $d=3$) 
\eq
F_2^{q}(R,L)=-\frac{\pi^2 L}{1152\ \sigma R^3}\left[2 E_4(\tau)-E_2^2(\tau)\right] \;\;.
\label{nlo}
\en

where $E_2$ and $E_4$ are the Eisenstein functions. They can be expressed in power series
as follows 
\eqa
\label{defe2}
E_2(\tau)&=&1-24\sum_{n=1}^\infty \sigma_1(n) q^n\\
\label{defe4}
E_4(\tau)&=&1+240\sum_{n=1}^\infty \sigma_3(n) q^n\\
q&\equiv& e^{2\pi i\tau} \;\;, \ena 
where $\sigma_1(n)$ and $\sigma_3(n)$ are, respectively, the sum of all divisors of $n$
and of their cubes (1 and $n$ are included in the sum).

Besides the corrections coming from the self-interaction, there is also a {\it boundary
  term} related to the boundary conditions of the string. In fact, due to the presence of
the Polyakov loops, the string has fixed ends and the effective string action may contain
terms localized at the boundary. The simplest possible term of this type is~\cite{lw02}:

\eq
\label{bterm}
\mathcal{A}_b=\frac{b}{4} \int_0^L \!\!\! dt
\left[  \left( \frac{\partial h}{\partial z} \right)_{z=0}^2
+  \left( \frac{\partial h}{\partial z} \right)_{z=R}^2
\right]
\en
where $b$ is a parameter with the dimensions of a length and $h$ denotes the transverse
displacement of the string; the factor $\frac{1}{4}$ has been added to agree with the
conventions of \cite{lw02}.  This additional term can be treated in the framework of the
zeta function regularization similarly to the computation of the free string correction.
The result of this calculation (see~\cite{chp04,chp04c} for details) is that, at first
perturbative order in $b$, the regularization of the free string action plus the boundary
term gives the same result as the pure free string action (namely the Dedekind function
discussed above), provided one replaces the distance $R$ between the quarks by:
\eq
R \to R^*=\frac{R}{(1+2\frac{b}{R})^{\frac12}}
\en
Thus, denoting by $F_b^q(R,L)$ the contribution to $F^q(R,L)$ due to the boundary term, we
have at first order in $b$ (remember that we have fixed $d=3$)
\eq
F_1^q(R,L)+F_b^q(R,L)=\log \eta \left( i \frac{L}{2R^*} \right)
\en
Exploiting again eq. (\ref{etafun}) and the modular transformation of the Dedekind $\eta$
function, this expression yields to the following behaviors
\begin{description}
\item $\bullet$ $2R^*<L$
\eq
F_1^q(R,L)+F_b^q(R,L)=-\frac{\pi L}{24 R}\left(1+\frac{b}{R}\right)
\label{eqbshort}
\en
\item $\bullet$ $2R^*>L$
\eq
F_1^q(R,L)+F_b^q(R,L)=-\frac{\pi R}{6 L}+\frac{\pi b}{6L}+
\frac{1}{2} \log\frac{2R}{L} - \frac{b}{2R}
\label{eq4}
\en
\end{description}
Besides the static quark potential defined in eq.~(\ref{statpot}), the
other observables we take into account are the force $Q(R)$
\eq
Q(\bar R)\equiv V(R+1) - V(R) = -\frac1L\log\left(\frac{G(R+1)}{G(R)}\right)
\label{defQ}
\en
and the combination~\cite{lw02} $c(R)$ related to the second derivative of $V(R)$
\eq
c(\widetilde R)\equiv -\frac12 \widetilde R^3 (V(R+1) + V(R-1) -2 V(R)) =
\frac12 \widetilde R^3\frac1L\log\left(\frac{G(R+1)G(R-1)}{G(R)^2}\right)~~~.
\label{defc}
\en 
In order not to enhance the lattice artifacts when considering discretized derivatives of
the static potential $V(R)$, in these definitions we have considered the following 
quantities~\cite{Som93,lw02}
\eq\label{Rbar}
\bar R (R) ^{-1} = 2\pi \left[ \Delta_2 (R-a) - \Delta_2 (R) \right]/a
\en
\eq\label{Rtilde}
\widetilde R (R) ^{-2} = 2\pi \left[ \Delta_2 (R+a) + \Delta_2 (R-a) -2 \Delta_2 (R) \right]/a^2
\en
where $\Delta_2 (R)$ is the Green function between the origin and the point $(R,0)$ of the
lattice Laplacian in 2 dimensions and $a$ is the lattice spacing. In table~\ref{RbatRtilde} 
we report the values of $\bar R(R)$ and $\widetilde R (R)$ up to $R=20$.
\begin{table}[ht]
\begin{center}
\begin{tabular}{|c|c|c|}
\hline
$R$ & $\bar{R}$ & $\widetilde{R}$\\
\hline
 3& 2.379& 2.808\\
 4& 3.407& 3.838\\
 5& 4.432& 4.875\\
 6& 5.448& 5.902\\
 7& 6.458& 6.920\\
 8& 7.464& 7.932\\
 9& 8.469& 8.941\\
10& 9.473& 9.948\\
11&10.475&10.953\\
12&11.478&11.957\\
13&12.480&12.961\\
14&13.481&13.964\\
15&14.483&14.966\\
16&15.484&15.968\\
17&16.485&16.970\\
18&17.486&17.972\\
19&18.486&18.973\\
20&19.487&19.975\\
\hline
\end{tabular}
\end{center}
\caption{Values of $\bar R(R)$ and $\widetilde R (R)$ (see eq.s~(\ref{Rbar})
  and~(\ref{Rtilde})) up to $R=20$.} 
\label{RbatRtilde}
\end{table}
Both quantities $Q(R)$ and $c(R)$ contain informations about the string effects in the static
quark potential. Since $Q(R)$ is related to the first derivative of $V(R)$, it is easier
to evaluate in numerical simulations but it depends explicitly on the string tension.
Thus, if the uncertainty in the determination of $\sigma$ is not small enough, it can affect
the subleading effects in which we are interested in.  On the contrary $c(R)$ only depends
on the contributions related to the quantum fluctuations of the flux tube and can
directly probe the reliability of the effective string description.  If the effective
string corrections are absent (like in the strong coupling phase below the roughening
transition) then $c(R)=0$. Hence, even if the numerical estimates for $c(R)$ are in general
less precise than those for $Q(R)$, in the following we shall mainly use $c(R)$ in our
analysis. The only exception will be the study of the long distance/high T data in
section \ref{sect5}, where we shall be able to extract important informations also from
$Q(R)$ by using the high precision estimate of $\sigma$ obtained at low temperature.

It is useful to write explicitly the predictions of the effective string description in the
short and large distance limits for $Q(R)$ and $c(R)$.

\noindent{\bf Short distance/low $T$ regime: $2R < L$}.\\
In the free bosonic string case, when $2R < L$ the Dedekind function can be approximated
to $-\pi L/24 R$ (i.e the L\"uscher term only), neglecting the exponentially decreasing
corrections. In this way, we obtain
\eq
Q_{1}(R)=\sigma-\frac{\pi}{24}(\frac{1}{R+1}-\frac{1}{R})=
\sigma+\frac{\pi}{24R(R+1)}
\en
and
\eq
c_{1}(R)=\frac{\widetilde R^3}{2}\frac{\pi}{24}(-\frac{2}{R}+\frac{1}{R+1}
+\frac{1}{R-1})=
\frac{\pi}{24}\frac{\widetilde R^3}{R(R^2-1)}
\en
Hence the free bosonic string action predicts that, as $R$ goes large (always fulfilling
the constraint $2R< L$), $c(R)$ approaches the asymptotic value
$\lim_{R\to\infty}c(R)=\frac{\pi}{24}$ which is the well known coefficient of the
L\"uscher term in 3-d. 

Let us now look at the corrections given by the Nambu-Goto string model truncated at the
second order. For $2R < L$, the Eisenstein functions can be approximated to 1 (see
eq.s (\ref{defe2}) and (\ref{defe4})~) up to exponentially small corrections and we obtain 
\eq
Q_{2}(R)=Q_{1}(R)-\frac{\pi^2}{1152\sigma}
(\frac{1}{(R+1)^3}-\frac{1}{R^3})
\label{eq:q2a}
\en
Similarly one finds:
\eq c_{2}(R)=c_{1}(R)+
\frac{\widetilde R^3}{2}\frac{\pi^2}{1152\sigma}(-\frac{2}{R^3}+\frac{1}{(R+1)^3} +\frac{1}{(R-1)^3})
\label{c2small}
\en
Finally, if a boundary term is present, we find an additional contribution to be added to
the previous ones 
\eq
Q_{b}(R)=\frac{b\pi}{24}\frac{2R+1}{R^2(R+1)^2}
\label{qbound}
\en
and
\eq
c_{b}(R)=\frac{b\pi}{12}\frac{\widetilde R^3}{(R^2-1)^2}
\label{cbound}
\en

\noindent{\bf Large distance/high $T$ regime: $2R > L$}.\\
Similarly to the short distance/low $T$ regime, we consider the free string
approximation and the Nambu-Goto model. In order to construct the expansions in the
limit $2R > L$, one performs a modular transformation of the Dedekind and Eisenstein
functions. For the Dedekind function this transformation is given in eq.(\ref{zbigtot}),
for the Eisenstein functions they have the following expressions
\eq
E_2(i\frac{L}{2R})=-\frac{4 R^2}{L^2} E_2(i\frac{2R}{L})+\frac{12R}{\pi L}\sim-\frac{4 R^2}{L^2}\\
\en
\eq
E_4(i\frac{L}{2R})= -\frac{16 R^4}{L^4} E_4(i\frac{2R}{L})+\frac{12R}{\pi L}\sim-\frac{16 R^4}{L^4}
\en
Neglecting again exponentially decreasing terms, we obtain the following expressions
(as above, the indices '1' and '2' refer to the free string approximation and to the
Nambu-Goto model respectively)
\eq
Q_{1}(R)=\sigma-\frac{\pi}{6L^2}+\frac{1}{2L}\log{(1+\frac1R)}
\en
\eq
Q_{2}(R)=Q_{1}(R)-\frac{\pi^2}{72L^4}-\frac{1}{8\sigma
L^2}\frac{1}{R(R+1)}
\en
\eq
Q_{b}(R)=\frac{b}{2L}\frac{1}{R(R+1)}
\en
and
\eq
c_{1}(R)=-\frac{\widetilde R^3}{4L}\log{(1-\frac{1}{R^2})}
\en
\eq
c_{2}(R)=c_{1}(R)-\frac{1}{8\sigma L^2}\frac{\widetilde R^3}{R(R^2-1)}
\en
\eq
c_{b}(R)=\frac{b}{2 L}\frac{\widetilde R^3}{R(R^2-1)}
\en
These are the expressions with which we shall compare our results in section \ref{sect5}.

It is useful to expand the previous expression in $1/R$ in order to better understand the
meaning of these terms
\eq
Q(R)\sim \left(\sigma-\frac{\pi}{6L^2}-\frac{\pi^2}{72L^4}\right)
+\frac{1}{2LR}+\frac{1}{R^2}\left(\frac{b}{2L}-\frac{1}{8\sigma
L^2}-\frac1{4L}\right)+\cdots
\label{qlargeR}
\en
and
\eq
c(R)\sim \frac{R}{4L}+\left(\frac{b}{2L}-\frac{1}{8\sigma L^2}\right)+\cdots
\label{clargeR}
\en
Eq. (\ref{qlargeR}) shows that the effective string description can also account for finite
temperature corrections to the zero temperature string tension. Eq. (\ref{clargeR})
instead predicts a linear dependence of $c(R)$ with the distance $R$.

\section{The model and the simulation parameters}
\label{sect2}
We have studied the (2+1) dimensional $SU(2)$ Yang-Mills theory on the
lattice with the Wilson action
\eq
S=\beta\sum_\Box (1-\frac12{\mbox{Tr}\; U_\Box})
\label{wilson}
\en
The sum is over all the plaquettes of a cubic lattice with $L$ spacings in the temporal
direction and $N_s$ in the two spatial ones. The gauge coupling is denoted by $\beta$ and
$U_\Box$ is the path ordered product of the gauge field along the plaquette. We impose
periodic boundary conditions in order to consider the Yang-Mills theory also at finite
temperature.

The Yang-Mills theory in (2+1) dimensions is not scale invariant at the classical level
since the continuum gauge coupling $g^2$ is a dimensionful quantity. It has the dimensions
of a mass and it is related to the dimensionless gauge coupling $\beta$ by
\eq
\frac{4}{g^2}=a\beta
\en
where $a$ is the lattice spacing. Hence, in (2+1) dimensions, scale invariance is
explicitly broken in the Lagrangian and it does not result -- as in (3+1) dimensions --
from the dynamical phenomenon of dimensional transmutation.

Close to the continuum limit all dimensionful quantities like the string tension at zero
temperature $\sigma(0)$ or the deconfinement temperature $T_c$ can be written as power
series in $a g^2/4=\beta$. The first few coefficients of these series are known quite
accurately~\cite{teper99}
\eq
a\sqrt{\sigma(0)} =\frac{1.324(12)}{\beta} +\frac{1.20(11)}{\beta^2} +.... 
\label{eq1}
\en
\eq
aT_c =\frac{1.50(2)}{\beta} +.... 
\en  

It is important to stress that the string tension $\sigma(T)$ depends on the temperature:
it decreases as the temperature increases and vanishes when $T\to T_c$. There are by now
clear numerical evidences that the deconfinement transition is second order. Then
the Svetitsky-Yaffe conjecture~\cite{Sve82} suggests that the universality class of this transition
is the same as that of the 2 dimensional Ising model. Thus the expected critical behaviour
of the string tension is $\sigma(T) \propto (1-T/T_c)^{2\nu}$ with $\nu=1$.  
Many numerical investigations have confirmed with high accuracy the validity of this
expectation.  In the following we shall mainly use the zero temperature value of the
string tension and we shall simply denote it by $\sigma$. 
  
Our numerical simulations have been carried out mostly at $\beta=9$, corresponding to the lattice
spacing $a\simeq 0.074 \it{fm}$. At this value of the gauge coupling the inverse deconfinement
temperature $1/T_c$ is about 6 in lattice units and the thickness of the flux tube
connecting the quark-quark pair roughly corresponds to $R_c\equiv \sqrt{1.5/\sigma}\sim
7.5$. Hence $R_c$ can be considered as the scale below which the effective string picture is no
longer expected to be valid. In order to investigate the scaling behaviour, we have also
performed a numerical simulation at $\beta =7.5$, corresponding to the lattice spacing
$a \simeq 0.09\it{fm}$.
In table~\ref{settings} we have summarized the simulation parameters considered in our
study. 
\begin{table}[ht]
\begin{center}
\begin{tabular}{|c|c|c|}
\hline
$\beta$ & $L$ & $N_s$ \\
\hline
9.0 & 8 & 120 \\
9.0 & 42 & 42 \\
9.0 & 48,54,60 & 36 \\
7.5 & 48 & 32\\
\hline
\end{tabular}
\end{center}
\caption 
{\sl Set of couplings $\beta$ and lattice sizes $N_s^2\times L$ we have considered in our
  study.}
\label{settings}
\end{table}

We have studied the model in two distinct regimes:
\begin{itemize}
\item First we have studied the short distance/low $T$ regime, considering
  lattices with a large temporal extent: we have chosen $\beta=9$, $L=42,48,54$ and 60. In order to
  keep small the finite temperature string corrections -- depending on the ratio $R/L$ --
  we have measured the static quark potential up to distances $R=13$. The lattice size
  $N_s$ in the spatial directions has then been fixed so as to make negligible the echo
  effects due to the periodic boundary conditions. We have set $N_s=42$ for $L=42$ and
  $N_s=36$ for the other three cases $L=48,54,60$. The chosen set of parameters leads to several
  simplifications in the effective string predictions, allowing to measure accurately the
  L\"uscher term and to detect possible higher order corrections.  Having results at four
  different values of $L$, we have also been able to study the dependence on $L$ of the
  various corrections. In this regime we have performed a simulation at $\beta=7.5$
  on a $32^2\times 48$ lattice to investigate the corrections to scaling.
   
\item 
  Second we have investigated the long distance/high $T$ regime: we have considered
  $L=8$, corresponding to a temperature of $3/4 T_c$.  In this case we have measured the
  static quark potential at much larger distances -- up to $R=20$ -- in order to
  reach values of the ratio $2R/L$ much larger than one. Since the temperature is high,
  the value of the string tension is quite small. We have carried out our simulations on a
  lattice with spatial extent $N_s=120$ in order to make the echo effects due to the
  periodic boundary conditions negligible at $R=20$.  This choice of the lattice size has
  allowed us to explore the ``modular transformed'' region of the effective string
  predictions. In this way a set of very stringent tests both on the leading correction
  and on the possible higher order terms have been possible.
\end{itemize}

\section{The algorithm}
\label{sect3}

In this section we describe the numerical techniques we have used in our Monte Carlo
simulations. Since the effective string picture is a description holding in the low-energy
regime, one has to measure the correlation function of pairs of Polyakov loops very far
apart. This is a hard numerical task due to the exponential decrease of the
correlation. The algorithm recently proposed by L\"uscher and Weisz \cite{lw01} has been a
breakthrough in the numerical estimation of correlation functions of Polyakov loops: it
yields to an exponential reduction of the error bars w.r.t. the numerical methods
previously used.

The idea of the algorithm -- in some respects similar to that of the multihit~\cite{Par83}
-- is to reduce the short wavelength fluctuations.  By freezing the Monte Carlo dynamics
of a suitably chosen set of lattice links, one splits the lattice into sublattices non
communicating among themselves. Then the observables are built up combining independent
measurements performed within every sublattice. Suppose that a measurement of an
observable ${\cal{O}}$ is obtained using the results of averages ${\cal{O}}_{sub}$
computed in ${\cal{N}}$ different sublattices. If $N$ sublattice measurements have been
carried out, the combination of the sublattice averages ${\cal{O}}_{sub}$ corresponds to
$(N)^{\cal{N}}$ measurements of ${\cal{O}}$. Hence, at the cost of about $N$ measurements
in the whole lattice, one gets an estimate of ${\cal{O}}$ as if $(N)^{\cal{N}}$
measurements would have been carried out. However, this estimate is biased by the
background field due to the frozen links. The next step is then to average over the
background field, computing biased estimates for many values of the background field.

Let us consider the correlation function of two Polyakov loops 
\beq\label{PPcor}
\langle P(\vec 0) P(\vec x )^* \rangle =
\frac{1}{{\cal{Z}}} \int \prod_{y,\mu} d U_{y,\mu} \;
\mbox{Tr} \left[ U_{(\vec 0,0),4}^{} \ldots 
U_{(\vec 0,L-1),4}^{}\right]
\mbox{Tr} \left[ U_{(\vec x,0),4}^{ \mbox{*}} \ldots 
U_{(\vec x,L-1),4}^{\mbox{*}} \right]
\; e^{-S[U]}
\eeq

We now split the lattice along the temporal direction into ${\cal{N}}=L/n_t$ sublattices
with a temporal thickness of $n_t$ lattice spacings. If we now keep fixed the set
$V^{(s)}_k$ of all the spatial links with time coordinates 
$k n_t$, $k=0,\ldots , ({\cal{N}}-1)$, then the dynamics of every sublattice depends
only on the background field of the two frozen time-slices that sandwich it along the
temporal direction. Hence the sublattices are totally independent among
themselves. Keeping this slicing in mind, we rewrite eq.~(\ref{PPcor}) as follows

\beq\label{PPcorLW}
\langle P(\vec 0) P(\vec x )^* \rangle =
\int \prod _k d U^{(s)}_k\;
\twolink_{(\vec 0,\vec x)}^{\alpha \gamma \beta  \delta }[V^{(s)}_0,V^{(s)}_1]
 \ldots  
\twolink_{(\vec 0,\vec x)}^{\epsilon \alpha \rho \beta } [V^{(s)}_{({\cal{N}}-1)},V^{(s)}_0]
\; {\cal{P}}[V^{(s)}_k]
\eeq

where $\twolink [V^{(s)}_i,V^{(s)}_j]$ is the following sublattice average with fixed
spatial links $V^{(s)}_i$ and $V^{(s)}_j$ at the two boundaries in the temporal direction

\beq\label{subav}
\twolink_{(\vec 0,\vec x)}^{\alpha \gamma \beta  \delta }
[V^{(s)}_i,V^{(s)}_j]
\equiv
\int \prod_{y,\mu} d U_{y,\mu} 
\left[ U_{(\vec 0,0),4}^{} \ldots 
U_{(\vec 0,n_t-1),4}^{}\right]_{\alpha \gamma}
\left[ U_{(\vec x,0),4}^{ \mbox{*}} \ldots 
U_{(\vec x,n_t-1),4}^{\mbox{*}} \right]_{\beta  \delta }
\frac{e^{-S[U;V^{(s)}_i,V^{(s)}_j]}}{{\cal{Z}}[V^{(s)}_i,V^{(s)}_j]} 
\eeq

The sublattice partition function with fixed temporal boundaries
${\cal{Z}}[V^{(s)}_i,V^{(s)}_j]$ is defined by

\beq\label{subpf}
{\cal{Z}}[V^{(s)}_i,V^{(s)}_j] \equiv
\int \prod_{y,\mu} d U_{y,\mu}
e^{-S[U;V^{(s)}_i,V^{(s)}_j]}
\eeq

where $S[U;V^{(s)}_i,V^{(s)}_j]$ is the Wilson action of the sublattice
with fixed temporal boundaries $V^{(s)}_i$ and $V^{(s)}_j$. In eq.~(\ref{PPcorLW}) 
$\alpha,\beta,\gamma,\delta$ are color indices and the contraction
rule is 
$\twolink^{\alpha \gamma \beta  \delta } \twolink^{\gamma \sigma
  \delta \tau } = \twolink^{\alpha \sigma \beta\tau }$. It is
important to note that $\twolink^{\alpha \gamma \beta  \delta }$ are
gauge-invariant quantities under sublattice gauge transformations. 
Finally the quantity ${\cal{P}}[V^{(s)}_k]$ is the probability for the
spatial links with time coordinates $k n_t$, $k=0,\ldots , ({\cal{N}}-1)$,
to be $V^{(s)}_k$

\beq\label{bgdist}
{\cal{P}}[V^{(s)}_k] =
\frac{1}{{\cal{Z}}}
\int \prod_{y,\mu} d U_{y,\mu} \;
\prod_k \delta( U^{(s)}_k - V^{(s)}_k) 
\; e^{-S[U]}
\eeq

Eq.~(\ref{PPcorLW}) states that if we evaluate $\twolink^{\alpha \gamma \beta \delta}$ in
every sublattice by performing $N$ sublattice measurements, then we obtain an estimate of
$\langle P(\vec 0) P(\vec x )^* \rangle$ -- at fixed $V^{(s)}$ -- as if $(N)^{\cal{N}}$
measurements would have been carried out. Then, changing the background field according
to the probability distribution of eq.~(\ref{bgdist}) and repeating the above procedure,
one completes the numerical computation of the integral of eq.~(\ref{PPcorLW}).  The
described technique is the so-called ``single level algorithm''. L\"uscher and Weisz have
also presented a generalized ``multilevel algorithm'' in which the updating frequency of
the background field $V^{(s)}_k$ is not the same for the various time slices $k$. However,
it seems that the ``single level algorithm'' is more efficient \cite{lw02} and this is
also what turns out from our experience.

When performing a numerical simulation with the single level algorithm, one has to fix the
value of 3 parameters: the temporal sublattice thickness $n_t$, the number $N$ of
sublattice measurements of $\twolink_{(\vec 0,\vec x)}$ and the number $M$ of background field
configurations to carry out the integration over $V^{(s)}$. These 3 parameters are related
among themselves and finding their optimal choice is a not trivial step. Moreover, they
also depend on the temporal extension $L$ of the whole lattice and on the distance
$R=|\vec x|$ between the two Polyakov loops. Some results on the optimization step can be
found in \cite{maj,Mey03}.

The temporal sublattice thickness $n_t$ has a minimal value coming from the knowledge we
have that the Polyakov loop correlation function decreases exponentially 
$\langle P(\vec 0) P(\vec x )^* \rangle \sim e^{-\sigma L R}$. If the sublattice temporal
extension $n_t$ is such that the sublattice is in the confined phase, then 
$\twolink_{(\vec 0,\vec x)}\sim e^{-\sigma n_t R}$. Hence we evaluate the small number 
$e^{-\sigma L R}$ as the product of ${\cal{N}}$ much larger numbers $e^{-\sigma n_t R}$. 
In this respect, the L\"uscher and Weisz algorithm exploits a similar solution the snake
algorithm \cite{deF00,Pep01} uses to compute accurately the exponentially small value of the 
't~Hooft loop.  An estimate of the sublattice minimal thickness is $L_{crit}/2$, where
$L_{crit}$ is the temporal extension at which the deconfinement transition takes place at
a given gauge coupling $\beta$.

The final error bar of the measure of $\langle P(\vec 0) P(\vec x)^*\rangle$ is the combination
of the uncertainties of the sublattice averages $\twolink_{(\vec 0,\vec x)}$ (depending on $N$)
and of the fluctuations of $\twolink_{(\vec 0,\vec x)}$ due to different background fields
(depending on $M$). If we suppose to have fixed $n_t$, the larger is $R$ the bigger must
be both $N$ and $M$. Typically, $N$ is order of several thousands and $M$ of few hundreds.
We note that $N$ does not depend on $L$ while $M$ does since it is related to the number
${\cal{N}}$ of frozen time-slices. According to our experience, when $2L/L_{crit}$
is large -- i.e. at very low temperature -- it is often better to consider sublattices
with a thickness $n_t$ bigger than the minimal one. Although $N$ must then be increased,
the reduction of the number of frozen time-slices ${\cal{N}}$ -- and, hence, also of $M$
-- makes the choice convenient. Indeed, in our short distance/low $T$ simulations at
$\beta =9$, we have considered $n_t=6$ even if the minimal thickness was $n_t=3$. However,
we have not performed a systematic study on the tuning of the parameters. 

As a last remark, we note that the L\"uscher and Weisz algorithm allows an exponential
gain in the accuracy of the numerical estimation of $\langle P(\vec 0) P(\vec x)^*\rangle$
only in the temporal direction. Indeed, every sublattice estimate 
$\twolink_{(\vec 0,\vec x)}\sim e^{-\sigma n_t R}$ is a quantity decreasing exponentially
fast with $R$ but it is still estimated by ``brute force'', i.e. with an error reduction
proportional to $1/\sqrt{N}$.  It is possible to devise a further, partial 
gain also in the measurement of the sublattice averages. Similarly to the above described
time slicing, one can cut every sublattice by freezing space-slices so that every spatial
direction is halved. Hence a $d+1$ dimensional sublattice will be split in $2^d$
independent sub-sublattices. Then, if the spatial positions, $\vec 0$ and $\vec x$, of the
two Polyakov loops are in different sub-sublattices, it follows that $K$ measurements
performed in each single sub-sublattice will result in $K^2$ measurements of the
correlator. In the numerical results we present in this paper, we have not implemented
this procedure for an additional partial gain: however it can become convenient when the
two Polyakov loops are very far apart.

\section{Analysis of the short distance/low $T$ data}
\label{sect4}
In this section we present the results we have obtained from the numerical simulations in
the short distance/low $T$ regime. The values that we obtained for $Q(R)$ and  $c(R)$ are
reported in tables~\ref{tabQ} and \ref{tabc} respectively.

\begin{table}[ht]
\begin{center}
\begin{tabular}{|c|c|c|c|c|}
\hline
$R$ & $L=42$ & $L=48$ & $L=54$ &$L=60$   \\
\hline
  2 &  0.040462(11) &0.0404558(87)&0.040437(10)&0.0404552(73)\\
  3 &  0.034381(14) &0.034373(11)&0.034349(13)&0.034374(10)\\
  4 &  0.031433(17) &0.031422(14)&0.031394(16)&0.031426(12)\\
  5 &  0.029790(20) &0.029778(16)&0.029747(19)&0.029785(14)\\
  6 &  0.028779(22) &0.028766(18)&0.028733(21)&0.028777(16)\\
  7 &  0.028112(25) &0.028099(20)&0.028063(24)&0.028113(17)\\
  8 &  0.027648(27) &0.027636(22)&0.027597(26)&0.027654(19)\\
  9 &  0.027312(29) &0.027302(23)&0.027260(29)&0.027322(22)\\
 10 &  0.027061(31) &0.027052(25)&0.027009(32)&0.027076(24)\\
 11 &  0.026869(34) &0.026862(28)&0.026819(35)&0.026887(28)\\
 12 &  0.026717(36) &0.026714(32)&0.026672(40)&0.026739(33)\\
 13 &  0.026597(40) &0.026600(38)&0.026554(48)&0.026616(40)\\
\hline
\end{tabular}
\end{center}
\caption{\sl Values of $Q(R)$ for various values of $L$ at $\beta=9$.}\label{tabQ}
\end{table}

\begin{table}[ht]
\begin{center}
\begin{tabular}{|c|c|c|c|c|}
\hline
$R$ & $L=42$ & $L=48$ & $L=54$ &$L=60$   \\
\hline
3&0.067293(40)&0.067319(32)&0.067382(39)&0.067299(28)\\
4&0.083360(93)&0.083418(74)&0.083527(91)&0.083345(66)\\
5&0.09519(18)&0.09527(14)&0.09543(17)&0.09508(13)\\
6&0.10392(30)&0.10402(24)&0.10427(30)&0.10367(24)\\
7&0.11058(48)&0.11062(39)&0.11104(50)&0.10996(39)\\
8&0.11583(73)&0.11558(63)&0.11629(81)&0.11465(64)\\
9&0.1198(11)&0.1195(10)&0.1205(13)&0.1185(11)\\
10&0.1236(16)&0.1226(16)&0.1233(21)&0.1212(18)\\
11&0.1266(23)&0.1252(28)&0.1250(35)&0.1244(31)\\
12&0.1298(34)&0.1262(51)&0.1256(62)&0.1262(56)\\
13&0.1310(54)&0.124(10)&0.128(12)&0.134(11)\\
\hline
\end{tabular}
\end{center}
\caption{\sl Values of $c(R)$ for various values of $L$ at $\beta=9$.}\label{tabc}
\end{table}

It is interesting to compare our data with those of ~\cite{maj} which were obtained in a
similar range of temperature and interquark distances. In particular, the sample which
allows the most natural comparison is the one at $\beta=10$ and $L=48$ reported in table~2
of \cite{maj}. While $Q(R)$ contains a residual scale dependence due to $\sigma$, the
function $c(R)$ can be directly compared. It is easy to see that our values are fully
compatible with the ones of~\cite{maj} and slightly more precise.

The data for $c(R)$ collected in table \ref{tabc} clearly show a smooth convergence
toward the expected free bosonic string value, $\frac{\pi}{24}=0.13089...$: the last two
values are already compatible with this result. Furthermore, it also turns out that the
data corresponding to different extensions $L$ in the temporal direction are compatible.
This indicates that -- within the errors bars -- not only the L\"uscher term, but also the
higher order corrections to the free energy are proportional to $L$ (so that they become
$L$ independent in $Q(R)$ and $c(R)$).

\begin{figure}[htb]
\centering
\includegraphics[height=10cm]{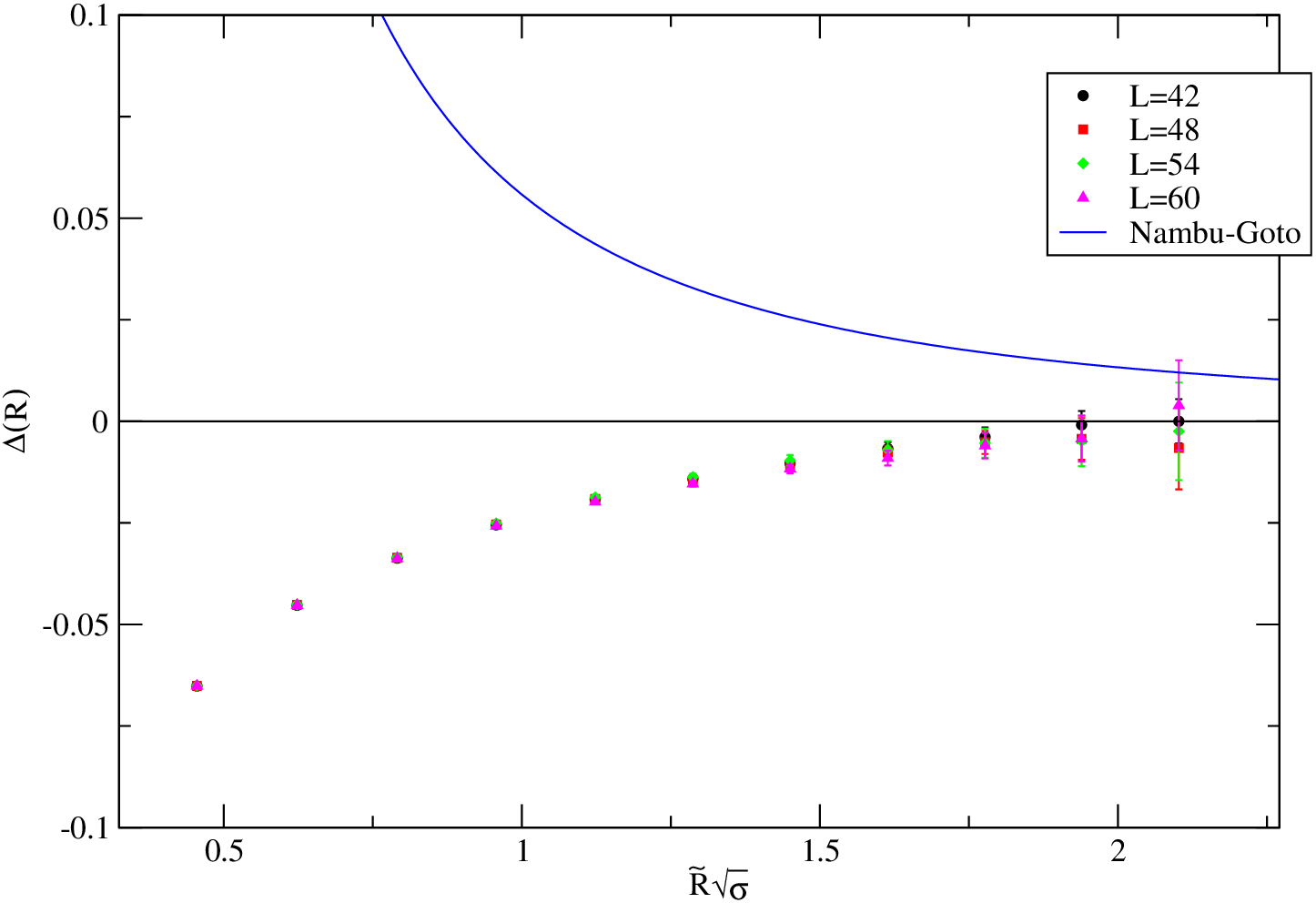}
\caption{Deviation of the results of our numerical simulations at $L=42$, 48,54, 60 and of
  the Nambu-Goto string model truncated at the second order (continuous line) from the
  free bosonic string approximation.} 
\label{figdelta}
\end{figure}

Based on these results, we have explored the higher order corrections studying the
deviations of the numerical data of $c(R)$ from the free bosonic string predictions of
eq. (\ref{bos}). Thus we have considered the following difference
\eq
\Delta(\widetilde R)\equiv c(\widetilde R)+\frac{\widetilde R^3}{2L}\left(F_1^q(R+1,L)+F_1^q(R-1,L)
-2F_1^q(R,L)\right)~~~.
\label{defdelta}
\en
The data for the four values of $L$ are plotted in figure~\ref{figdelta}. As anticipated
above, it clearly turns out that the data for 
different $L$ nicely agree among themselves. It is important to address the question of the
range of interquark distances in which we must expect these higher order
corrections. As mentioned in section~\ref{sect2}, there is a natural scale~\cite{chp03},
$R_c$, below which one cannot trust any kind of effective string description. In fact at
this scale the interquark distance becomes comparable with the string width: the
internal structure of the flux tube becomes relevant and the approximation with a string
is no longer valid. Below this scale, one actually enters in the perturbative domain of the
theory (see the discussion in~\cite{lw02} and~\cite{maj}). At the value of the gauge
coupling we have used in our numerical simulations, we have $R_c\sim 7.5$ and so
we must consider only distances $R\geq 8$. Moreover, as discussed above, the numerical
data at $R=12,13$ are already compatible with the free bosonic string approximation and so
the higher order corrections are zero within the errors. Thus we concentrate our analysis
on the four values $R=8,9,10,11$. 
Our first goal is to estimate the dependence on $R$ of the correction to the free bosonic
string behaviour. To this end the simplest option would be to fit the data.  However with
this approach one has to face a major objection.  Due to the numerical method we use, data
at different distances $R$ are correlated and the crosscorrelation matrix is so flat that
keeping it into account in the fits is very difficult. To circumvent these problems we
decided to {\sl assume} a given correction, say $b_{\alpha}/R^\alpha$, extract the coefficient
$b_{\alpha}$ from each one of the four data at our disposal and then check the stability
of the results obtained in this way.  Let us see this in more detail.  Let us assume that
the potential is described by
\eq
V(R)\equiv -\frac1L G(R)= \sigma R +k +\frac{F_1^q(R,L)}{L}
+\frac{b_\alpha}{R^\alpha}
\en
then it is easy to see that $\Delta(R)$ should behave as 
\eq
\Delta(R)=- b_\alpha \frac{\widetilde R^3}{2}\left(\frac{1}{(R+1)^\alpha}+
\frac{1}{(R-1)^\alpha}-\frac{2}{R^\alpha}\right)
\label{eq0}
\en
From this equation one can estimate the value of the coefficient $b_\alpha$. It is
interesting to see the physical meaning of this parameter. A typical string
self-interaction term is modeled by the insertion in the Lagrangian of a quartic
interaction term (as in the case of the Nambu-Goto action) and leads (see for
instance~\cite{df83} or~\cite{chp03}) to a correction with $\alpha=3$. Instead a boundary
term would lead to a correction with $\alpha=2$. In particular, with our choice of
notations, we have $b_2=-\frac{\pi b}{24}$ (see eq.(\ref{eqbshort})~). 

As an example of the results of our analysis we report in table~\ref{tabb1} the values of
$b_\alpha$ for $\alpha=2$ and 3 for the data collected at $L=42$. For the case $\alpha=3$
the estimates at the various distances $R$ turn out to agree within the errors. The
results obtained using the data collected with the other values of $L$ are fully
compatible with those reported in table~\ref{tabb1}. Combining together the results
collected at the four different values of $L$ (they are completely uncorrelated since they
have been obtained from independent simulations), we obtain as our final estimate for $b_3$
the value $b_3= 0.145(15)$

\begin{table}[ht]
\begin{center}
\begin{tabular}{|c|c|c|}
\hline
$R$ & $b_2$ & $b_3$ \\
\hline
 8& 0.038(2) & 0.15(1) \\
 9& 0.032(3) & 0.14(1) \\
 10& 0.022(5) & 0.11(3) \\
 11& 0.014(9) & 0.08(5) \\
\hline
\end{tabular}
\end{center}
\caption{\sl Values of $b_2$ and $b_3$ from eq.(\ref{eq0}) using the data collected with $L=42$.} 
\label{tabb1}
\end{table}

Since the parameters $b_\alpha$ are dimensionful quantities, it is useful to study the
dimensionless ratios 
\eq
\gamma_\alpha\equiv b_\alpha\sigma^{(\alpha-1)/2} 
\en
This will allow us to compare our results with other pure gauge theories, with other
$SU(2)$ simulations performed at different couplings $\beta$ or with a theoretical model
like the Nambu-Goto one. For instance, in the $\alpha=3$ case, we find:
\eq
\gamma_3=0.0038(4)
\en

Unfortunately, the data do not allow to disentangle unambiguously a boundary term
contribution from the one due to a quartic correction in the string action. However the
stability of the numerical results for $b_3$ suggests that -- similarly to the $SU(3)$
case in $(2+1)$ dimensions -- there is no boundary term. Nevertheless, if we try to
interpret the deviations from the free bosonic string as a boundary correction, we find
that the boundary parameter $b$ must be negative and rather small, ranging from 
$b\sim -0.3$ for $R=8$ to $b\sim -0.1$ for $R=11$. Further investigations are needed to give more
definite statements in this respect.

Instead, if we assume that the boundary term is absent and that the correction is completely due to  a
selfinteraction term in the effective string action, then from our analysis
we obtain two important results which we present here and shall discuss
in greater detail in section~\ref{sect6}.
\begin{description}
\item{1]} The value of $\gamma_3$ that we find is similar (but not equal within the
  errors) to the one obtained in the $\Z (2)$ gauge theory at $\beta=0.73107$:
  $\gamma_{3,\Z(2)}=0.0047(1)$~\cite{chp04}.
\item{2]} The value of $\gamma_3$ that we obtain is definitely different from the
  Nambu-Goto expectation $\gamma_{3,NG}\equiv -\frac{\pi^2}{1152}=-0.00857.. $ (see
  eq.(\ref{nlo})~) which turns out to be opposite in sign and more or less double in
  magnitude. It is important to stress that a similar disagreement in
the short distance regime was recently reported
in~\cite{jkm02,jkm03,chp04}.
\end{description}

\subsection{Estimate of $\sigma$}\label{sec:sigma}
Our short distance/low $T$ results allow an estimate for the string tension which
turns out to be much more accurate than the existing ones at the considered value of
$\beta$.  Again, as discussed above, we cannot use a simple minded fit to obtain $\sigma$
from our data. We have chosen to use the results of the previous analysis to extract, 
{\sl for each value of $R$}, an estimate of the string tension from the measurements of
$Q(R)$. The stability of the extracted values as a function of $R$ also gives us a test of
reliability. Based on the previous analysis of the corrections to the free bosonic string,
we have assumed no boundary term and we have estimated $\sigma$ as
\eq
\sigma=Q(R)+\frac{\pi}{24}\left(\frac{1}{R+1}-\frac{1}{R}\right)-b_3
 \left(\frac{1}{(R+1)^3}-\frac{1}{R^3}\right)
\label{eqsigma}
\en
using $b_3=0.145$.  The results are reported in table~\ref{tabsigma}. Interestingly, we
note that, for each one of the four values of $L$, the data are compatible within the
errors for all the values $R>R_c$, i.e. for $R\geq 8$.  
For each $L$ we consider the $R=8$ result as our best estimate for $\sigma$ since it is the
one with the smallest uncertainty. The four values of $L$ are independent
experiments and we can average over them yielding the final estimate $\sigma=0.025900(12)$. It is
interesting to compare this result with the existing ones. In ref.~\cite{teper99} we find two
different {\sl direct} estimates obtained from two different lattice sizes:
$\sigma=0.02611(19)$ for $L=24$ and $\sigma=0.02631(13)$ for $L=32$. Using the
perturbative formula (\ref{eq1}) one 
obtains a value inbetween, but with a much larger uncertainty $\sigma=0.02622(45)$. Our
result is compatible with this last estimate, but it is definitely smaller than the two
direct estimates.  Finally it is also interesting to observe that the same pattern seems
to be present in the results of~\cite{maj} at $\beta=10$. At this value of $\beta$,
eq.(\ref{eq1}) gives $\sigma=0.02085(37)$ while ref.\cite{maj} obtains $\sigma=0.02065(6)$
which, as in our case, is smaller than the extrapolated one even if compatible within the
error bars. 

\begin{table}[ht]
\begin{center}
\begin{tabular}{|c|c|c|c|c|}
\hline
$R$ & $L=42$ & $L=48$ & $L=54$ &$L=60$   \\
\hline
2&0.031400(11)&0.0313938(87)&0.0313751(10)&0.0313932(73)\\
3&0.026578(14)&0.026569(11)&0.026545(13)&0.0265706(96)\\
4&0.025993(17)&0.025983(14)&0.025955(16)&0.025987(12)\\
5&0.025915(20)&0.025904(16)&0.025873(19)&0.025911(14)\\
6&0.025911(22)&0.025898(18)&0.025865(21)&0.025909(16)\\
7&0.025914(25)&0.025901(20)&0.025865(24)&0.025915(17)\\
8&0.025914(27)&0.025902(22)&0.025863(26)&0.025920(19)\\
9&0.025912(29)&0.025901(23)&0.025859(29)&0.025922(22)\\
10&0.025907(31)&0.025899(25)&0.025855(32)&0.025922(24)\\
11&0.025902(34)&0.025895(28)&0.025852(35)&0.025920(28)\\
12&0.025896(36)&0.025893(32)&0.025851(40)&0.025918(33)\\
13&0.025890(40)&0.025894(38)&0.025848(48)&0.025910(40)\\
\hline
\end{tabular}
\end{center}
\caption{\sl Values of $\sigma$ extracted using eq.(\ref{eqsigma}) and the data
of table~\ref{tabQ} as input.}\label{tabsigma}
\end{table}
\subsection{Scaling behaviour of $\gamma_3$}\label{sec:scaling}
\begin{figure}[htb]
\centering
\includegraphics[height=10cm]{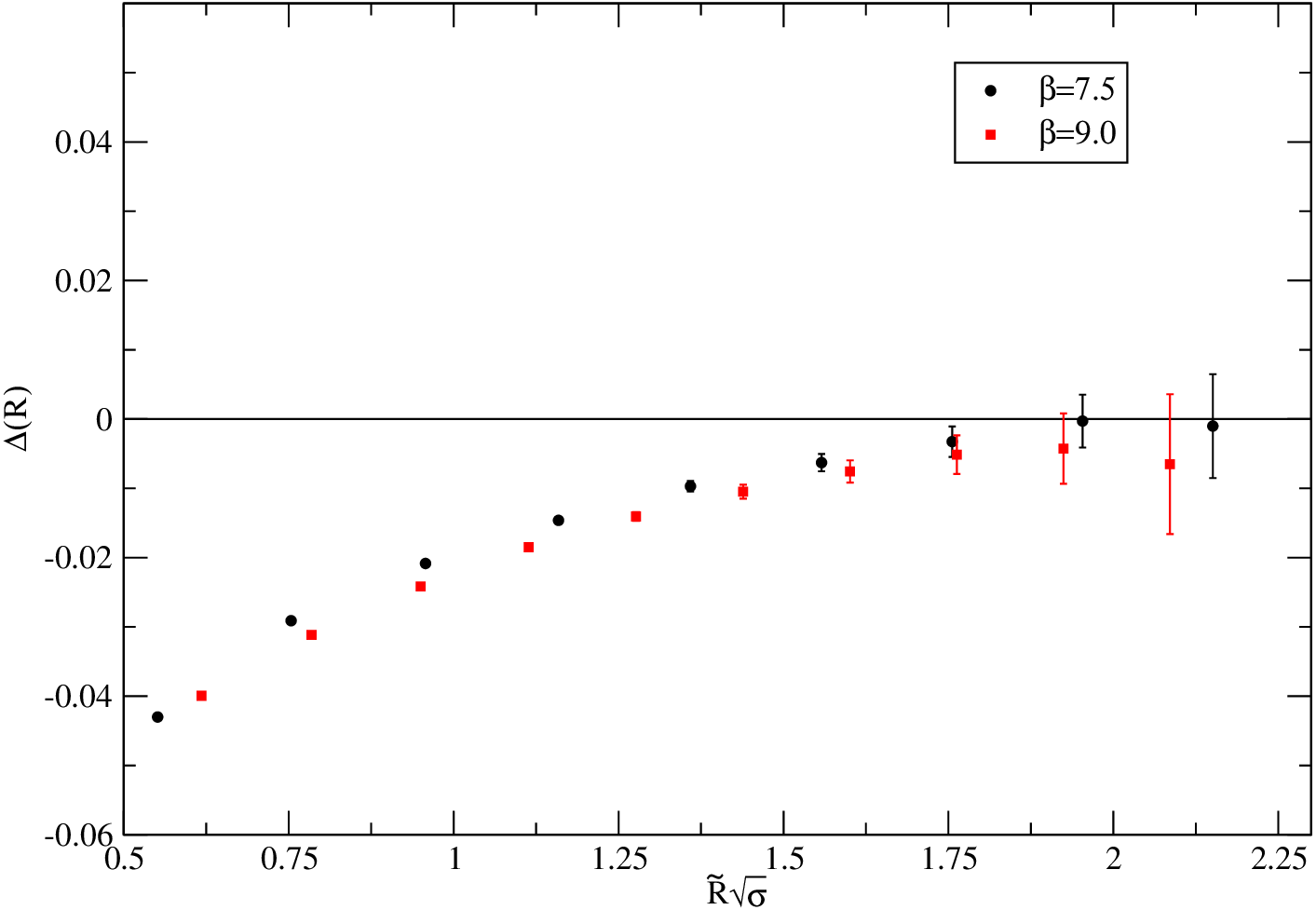}
\caption{Comparison of the results for $\Delta(R)$ as a function of the scaling variable 
$\widetilde R\sqrt{\sigma}$ for the two samples:
$\beta=9,~ L=48$ and  $\beta=7.5,~ L=48$}
\label{figscal}
\end{figure}

In order to test the scaling behaviour of the corrections we have found at $\beta=9$,
we have performed the same analysis on a data set collected at $\beta=7.5$ on a
$32^2\times 48$ lattice. We have observed the same behaviour as at $\beta=9$ and --
including in this case also the values at $R=6$ and 7 since $R_c =5$ -- we have obtained 
\eq
b_3=0.08(2)  ~~~~~\sigma=0.03856(4)
\en
From these data we estimate $\gamma_3=0.0031(7)$. This value is compatible with the one
obtained at $\beta=9$, indicating that the observed corrections are not due to scaling
violations\footnote{A similar analysis of possible scaling violations was performed
in~\cite{chp04} for the $\Z (2)$ gauge theory: the result is consistent with what we
observe here.}. This is well exemplified in the figure~\ref{figscal} where the data for
$\Delta(R)$ for the two samples ($\beta=9,\, L=48$ and  $\beta=7.5,\, L=48$) are compared
as a function of the scaling variable $\widetilde R\sqrt{\sigma}$. In table~\ref{databeta7.5} we report
the results for $Q(R)$ and $c(R)$ we have obtained at $\beta=7.5$.

\begin{table}[ht]
\begin{center}
\begin{tabular}{|c|c|c|}
\hline
$R$ & $Q(R)$ & $c(R)$\\
\hline
2&0.054827(13)& --- \\
3&0.047808(17)&0.077682(44)\\
4&0.044476(20)&0.094185(99)\\
5&0.042658(22)&0.10532(19)\\
6&0.041559(25)&0.11301(33)\\
7&0.040845(27)&0.11832(56)\\
8&0.040357(30)&0.12201(94)\\
9&0.040008(32)&0.1247(16)\\
10&0.039746(36)&0.1288(28)\\
11&0.039550(41)&0.1288(54)\\
12&0.039404(49)&0.125(11)\\
13&0.039304(66)&0.109(28)\\
\hline
\end{tabular}
\end{center}
\caption{Values of $Q(R)$ and $c(R)$ for the simulation at $\beta=7.5$ on a $32^2\times
  48$ lattice.}\label{databeta7.5}
\end{table}

\section{Analysis of the long distance/high $T$ data}
\label{sect5}
In this section we discuss the results we have obtained in the long distance/high $T$
regime. Our numerical simulations have been performed on a lattice with temporal extent
$L=8$. The values we have obtained for $Q(R)$ and $c(R)$ are reported in table~\ref{tabQc}.

\begin{table}[ht]
\begin{center}
\begin{tabular}{|c|c|c|}
\hline
$R$ & $Q(R)$ & $c(R)$\\
\hline
2&0.037412(41)& --- \\
3&0.030931(49)&0.071497(73)\\
4&0.027558(57)&0.09488(18)\\
5&0.025497(65)&0.11847(35)\\
6&0.024094(72)&0.14288(60)\\
7&0.023066(80)&0.16832(94)\\
8&0.022274(87)&0.1948(14)\\
9&0.021643(94)&0.2222(20)\\
10&0.02113(10)&0.2505(27)\\
11&0.02069(11)&0.2792(36)\\
12&0.02032(11)&0.3080(47)\\
13&0.02000(12)&0.3380(61)\\
14&0.01973(13)&0.3683(76)\\
15&0.01948(14)&0.3981(94)\\
16&0.01926(15)&0.430(11)\\
17&0.01905(15)&0.459(14)\\
18&0.01887(16)&0.491(17)\\
19&0.01869(17)&0.526(20)\\
\hline
\end{tabular}
\end{center}
\caption{\sl Values of $Q(R)$ and $c(R)$ for $L=8$ at $\beta =9$.}\label{tabQc}
\end{table}
\begin{figure}[htb]
\centering
\includegraphics[height=10cm]{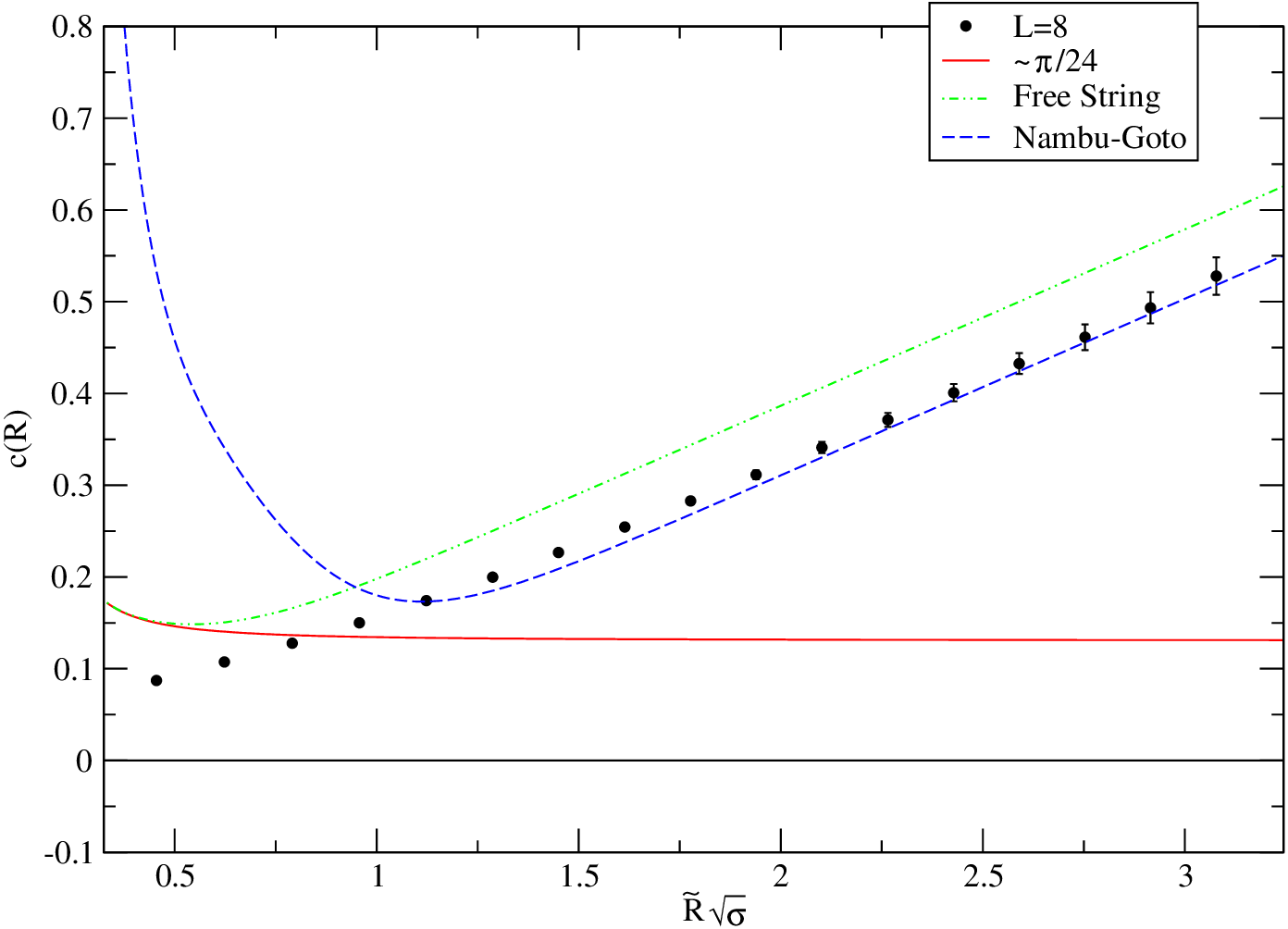}
\caption{Comparison of the numerical results for $c(R)$ at $L=8$, $\beta =9$, with the
  expectations for the free bosonic string (dashed-dotted line) and with the Nambu-Goto
  string model truncated at the second order (dashed line). The continuous line is what one
  would find assuming a correction to the linear behaviour like the L\"uscher term $\pi /24 R$.}
\label{figc8}
\end{figure}
\begin{figure}[htb]
\centering
\includegraphics[height=10cm]{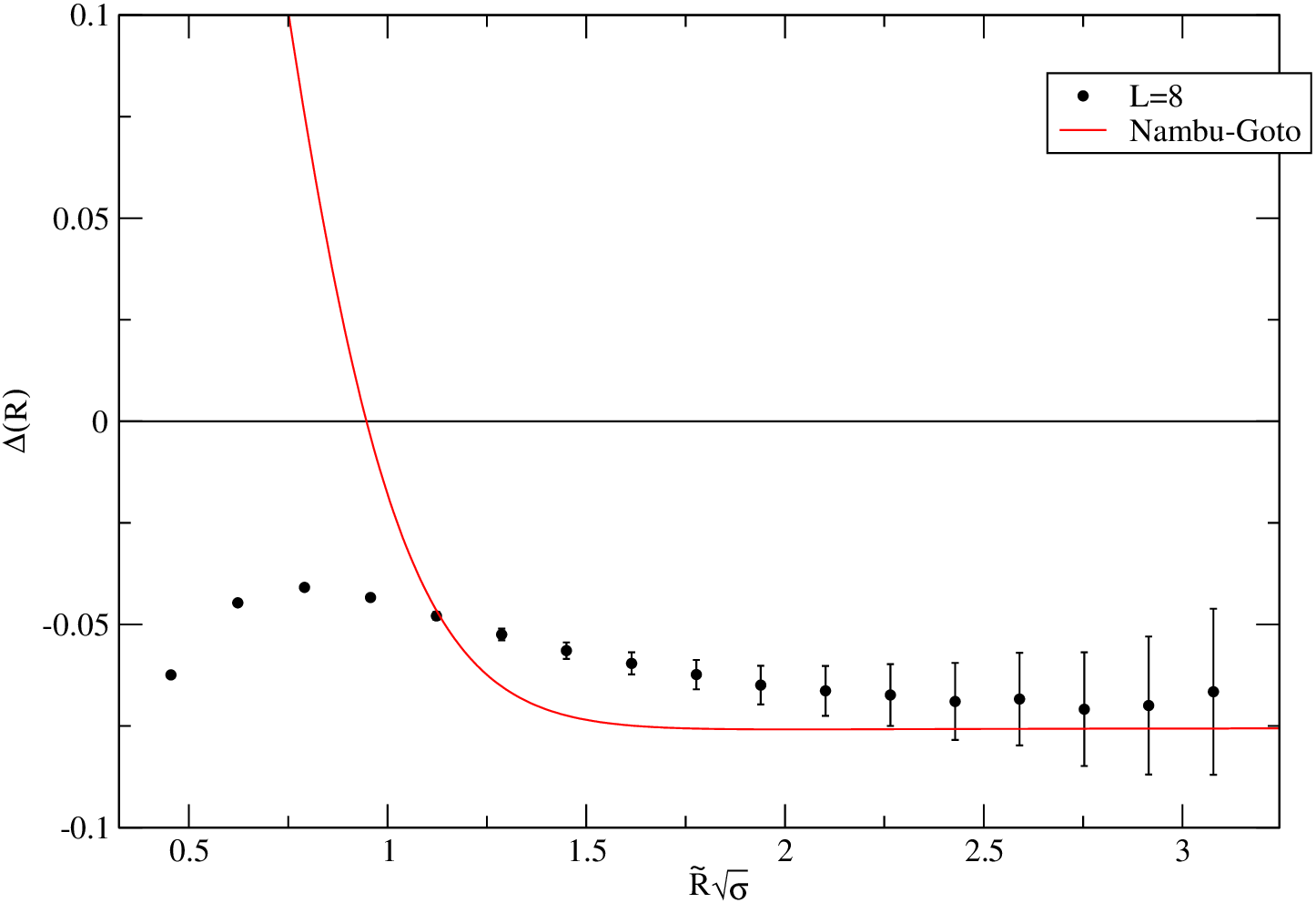}
\caption{Deviation of the results of our numerical simulation at $L=8$ and of
  the Nambu-Goto string model truncated at the second order (continuous line) from the
  free bosonic string approximation.} 
\label{figdelta8}
\end{figure}
\begin{figure}[htb]
\centering
\includegraphics[height=10cm]{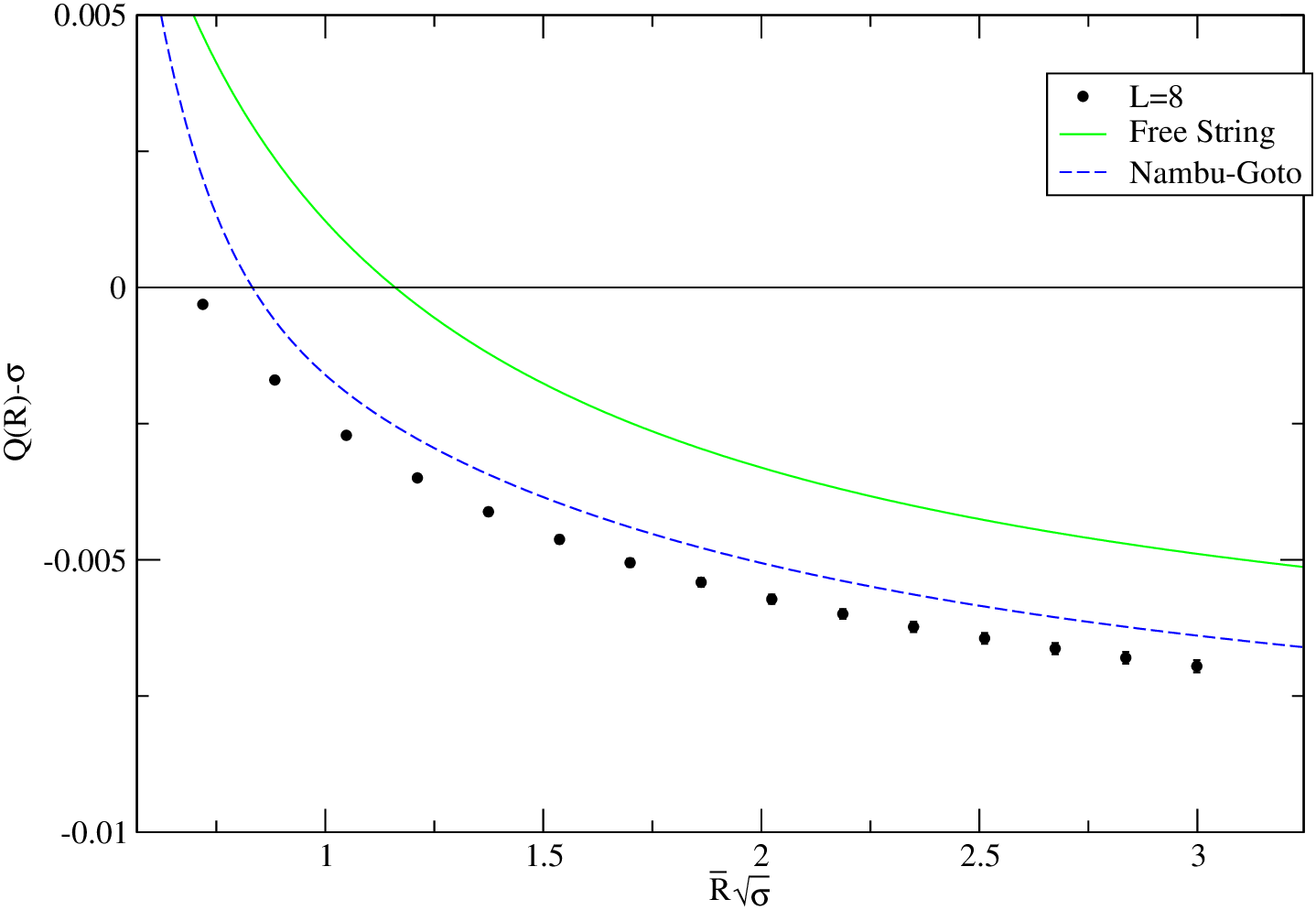}
\caption{Comparison of the numerical results for $(Q(R)-\sigma)$ at $L=8$, $\beta =9$, with the
expectations for the free bosonic string (continuous line) and for the Nambu-Goto string
truncated at the second order (dashed line).}
\label{figqdelta8}
\end{figure}

These data show that in the short distance/high $T$ regime the behaviour
of $c(R)$ is completely different from the one observed in the long distance/low $T$ case.
There is no evidence of a convergence toward the coefficient of the L\"uscher term in 3
dimensions, $\pi/24$, but instead a linearly rising behaviour is observed (see figure~\ref{figc8}). The
reason is that most of the data are in the $2R>L$ regime where a linearly rising
behaviour for $c(R)$ is exactly what one expects from the free bosonic effective string model
eq.~(\ref{clargeR}). Let us stress that this agreement is highly non trivial and
represents a very important test of the effective string description for the static quark
potential.  As a matter of fact, since eq.~(\ref{clargeR}) is obtained by modular
transforming the Dedekind function, its agreement with the numerical data is actually a
test of the whole functional form of the effective string correction\footnote{The
same holds also for the Eisenstein functions and thus for the Nambu-Goto contribution.}.
In other words the agreement is obtained only keeping into account all the exponentially
decreasing terms contained in the Dedekind function (which are mandatory ingredients in
the modular transformation). Thus, even if in a rather indirect way, it could be
considered as a test of the fact that the spectrum of the model in the short distance/low
$T$ regime should be the one predicted by the bosonic effective string model. 
This seems to disagree with the results recently discussed in~\cite{jkm03}, where the
authors investigate the spectrum of the stable excitations of the confining flux
tube. From this study it turns out that below 1$fm$ -- i.e. for a range of distances
where the L\"uscher term is already observed in the ground state of the static quark
potential -- the spectrum of the flux tube excitations is instead grossly distorted
w.r.t. the expectations based on the effective string description. Further studies are
needed to better understand  the connection between our indirect test and the direct
numerical evidence of~\cite{jkm03}.

Following the above comment, it is also important to notice that in the long distance/high
$T$ regime any attempt to fit the data with a ``L\"uscher''-type correction 
$c(R)\sim \pi/24$ leads to a complete disagreement with the numerical simulations (see the
continuous line in figure \ref{figc8}) already for values of $R$ comparable with the inverse
temperature. This is a rather important remark, since this fitting procedure is very
common in the literature and, in this regime, could lead to an apparent and erroneous
rejection of the effective string model or to a systematic error in the estimate of $\sigma$.
  
The difference between the free bosonic string and the Nambu-Goto model is, in general,
very small, but it increases as the temperature increases. The choice of $L=8$ was
motivated exactly by this consideration since at this temperature the difference between
the two effective string predictions is definitively larger than the errors of our
simulation and we can discriminate between them. As figure \ref{figdelta8} shows, the
situation in the long distance/high $T$ regime is completely different from the short
distance/low $T$ one discussed in the previous section.  In fact $\Delta(R)$ has now the
same sign of the Nambu-Goto prediction and its shape and numerical value are in good
agreement with it. This is one of our main results and suggests that, contrary to what
happens in the short distance/low $T$ regime, the Nambu-Goto string can give a reliable
description for the static quark potential in the large distance/high $T$ regime.
Interestingly, a similar scenario also occurs in the 3-d $\Z (2)$ gauge
theory~\cite{chp03}.

In principle one could interpret the deviation from the free bosonic string behaviour in
figure \ref{figdelta8} as due to a boundary term instead of a Nambu-Goto like correction.
This requires $b$ to have a value $b\sim -0.8$ which agrees in sign with the one that we
find under the same assumptions in the short distance/low $T$ regime, but it is from three
to eight times bigger. Again, this could suggest that the deviations in $\Delta(R)$ might
not be due to a boundary term. As a matter of fact the value of $b$ could be fixed without
ambiguities looking at different values of $L$ since the boundary correction and the
Nambu-Goto one have a different $L$ dependence. However this is not easy since it would
require to keep the high precision of our results also for larger values of $L$. This type
of analysis was indeed performed in the case of the 3-d $\Z(2)$ gauge theory~\cite{chp04}
where, thanks to the different nature of the simulation algorithm~\cite{deF00,Pep01,chp03}, this
level of precision can be reached for arbitrary values of $L$ and $R$. We plan to address
this type of analysis for the $SU(2)$ model in a forthcoming publication.
  
Further interesting informations can be obtained from $Q(R)$ (see
figure~\ref{figqdelta8}). As we mentioned in section~\ref{sect1}, the behaviour of $Q(R)$ is 
dominated by $\sigma$. However in the present case we can use the very precise value for sigma
obtained in section~\ref{sect4}. In figure~\ref{figqdelta8} we plot the difference
$(Q(R)-\sigma)$ as a function of $R$. Thanks to the accurate estimate of $\sigma$, the
systematic uncertainty in this difference due to $\sigma$ turns out to be negligible
w.r.t. the statistical errors of $Q(R)$. Figure~\ref{figqdelta8} shows that the numerical
estimates lie below the Nambu-Goto prediction truncated at the second order. On one
hand, this means that the Nambu-Goto prediction truncated at the second order describes
the data better than the free bosonic string, on the other hand it also suggests that the
higher order terms in the expansion of the Nambu-Goto action could fill the remaining gap
between numerical data and theoretical estimates. 
Finally let us notice that a similar agreement with the Nambu-Goto model was also
recently observed in the large distance/high $T$ data for the 3-d $\Z(2)$ gauge theory.

\section{Comparison with $\Z(2)$ and $SU(3)$ gauge theories}
\label{sect6}
It is very interesting to compare our short distance/low $T$ results with those recently
obtained for the $\Z(2)$~\cite{cpp02,chp03,chp04} and $SU(3)$~\cite{lw02} gauge theories.
Among the $SU(3)$ data published in~\cite{lw02} we have selected, for our comparison, the
sample at $\beta=20$ (see the data in table 3 of ref.~\cite{lw02}) which is the one with
the larger set of data.  For the $\Z (2)$ gauge theory we have chosen the sample at $\beta=0.73107$ and
$L=80$ of ref.\cite{chp04} which corresponds to a very low temperature $T/T_c=1/20$ and to
a similar value for the lattice spacing. For the $SU(2)$ case we have used for the comparison
only the $L=42$ sample which is the most precise one. The other three samples would add no
further information since they are compatible within the errors.  For distances $R$ large
enough, the three theories show the same asymptotic behaviour. Our aim is to study the
higher order corrections which characterize the approach to the asymptotic free bosonic
string limit. 
 
\begin{figure}
\centering
\includegraphics[height=10cm]{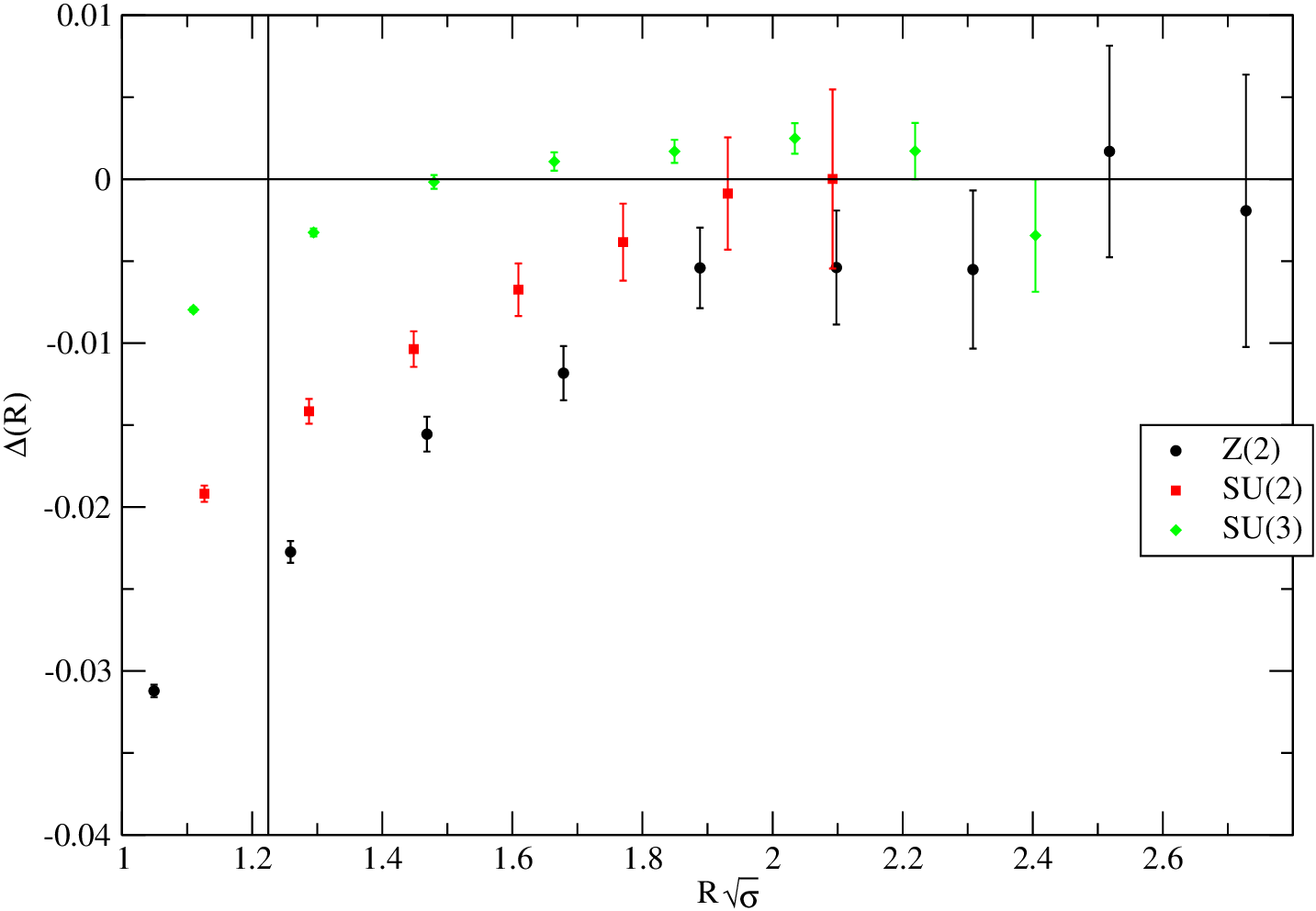}
\caption{Values of $\Delta (R)$ in 3-d for the $SU(3)$ Yang-Mills theory (diamonds), the
  $\Z (2)$ gauge theory (circles) and the sample with $L=42$ of the $SU(2)$
  Yang-Mills theory (squares). The vertical line corresponds to
  $R=R_c=\sqrt{1.5/\sigma}$. Here $\Delta (R)$ is evaluated considering $R$ and not
  $\widetilde R$.}
\label{su2comparison}
\end{figure}

We plot the three data sets in figure~\ref{su2comparison}. Since the three sets 
correspond to different values of $\sigma$, we have rescaled $R$ in units of
$\sqrt{\sigma}$ in order to make a meaningful comparison.

Let us briefly comment on figure~\ref{su2comparison}.
\begin{itemize}
\item As discussed in \cite{lw02}, the $SU(3)$ data are rather well described by the
  free string correction only and $\Delta(R)$ turns out to be very small for $R>R_c$.
  
\item On the contrary, both the $SU(2)$ and the $\Z (2)$ data show
  rather strong deviations from the free string behaviour. These deviations are stronger
  in the $\Z (2)$ case than in the $SU(2)$ one.  
  
\item For all the three models the observed corrections definitely disagree with the
  Nambu-Goto prediction (which is not reported in the figure but it can be seen in
  figure~\ref{figdelta}) which has the opposite sign and is larger in
  magnitude.

\end{itemize}

Our conclusion is that, in 3 dimensions, the terms subleading to the free bosonic string
behaviour in $\Z (2)$, $SU(2)$ and $SU(3)$ gauge theories do not show any evidence of 
common features. This confirms the expectation that these subleading corrections depend on
the string dynamics and are thus different for the different gauge groups.

A similar comparison between the 3-d $SU(2)$ and $\Z(2)$ gauge theories for the ground
state energy and for the first excited string level was recently performed in~\cite{jkm03,Jug04}.
In that case the two models were shown to have similar behaviours after readjusting the
ratio of the string tension to the glueball mass in the $\Z(2)$ gauge theory. It would be
interesting to understand if this agreement is a consequence of the rescaling or if it is
instead related to the fact that the two models share the same center. This is one of the
reasons for which we plan to extend our analysis also to the gauge theory with gauge group
$Sp(2)$~\cite{Hol03,Hol03b} which has the same center as $SU(2)$ and $\Z(2)$.

\section{Concluding remarks}
\label{sect7}

In this paper we have studied the corrections to the linear behaviour of the static quark
potential in the $SU(2)$ Yang-Mills theory in (2+1) dimensions. We have considered both
the zero and the finite temperature regimes comparing the numerical results with the
expectations coming from effective string descriptions of the quark interaction.
Finally we have checked for common features among the effective string descriptions of 
$\Z(2)$, $SU(2)$ and $SU(3)$ gauge theories. Our results can be summarized as follows.

\begin{itemize}
\item Our numerical simulations strongly support the conjecture that the ground state of
  the static quark potential in the (2+1) dimensional $SU(2)$ Yang-Mills theory (as well
  as in all the other pure gauge theories studied up to now) is very well described by an
  effective string theory. Apart from subleading corrections -- which can be observed at
  short distances and/or at high $T$ -- this effective string theory is the
  bosonic string theory originally suggested by L\"uscher, Symanzik and Weisz in~\cite{lsw}.
 
\item We observe deviations from the free string behaviour also in the short distance/low $T$
  regime.  These deviations definitely disagree with the Nambu-Goto predictions.
  A similar disagreement was recently observed also in the 3-d $\Z (2)$ gauge
  theory~\cite{jkm02,jkm03,chp04}. 
 
\item In the large distance/high $T$ regime, we clearly observe deviations from the free
  bosonic string approximation. These deviations qualitatively -- and, in part, also
  quantitatively -- agree with the predictions of the Nambu-Goto action truncated at the
  second perturbative order. 
   
\item The large distance/high $T$ regime can be obtained by a modular transformation
  of the effective string results in the short distance/low $T$ regime. The observed
  agreement with the effective string expectation not only confirms the presence of the
  L\"uscher term but also suggests an indirect check of the whole functional form and, in
  particular, of the spectrum of the excited states. However we emphasize that this is
  still an open question that deserves further investigations.
  
\item Comparing the $\Z (2)$, $SU(2)$ and $SU(3)$ gauge theories, we do not find any clue
  for common features in the corrections to the L\"uscher term among them. 
\end{itemize}

In view of this last point, it would be important to understand if these analogies and
differences are related to the gauge group, to its center, or are non-universal features,
perhaps related to different values of the boundary term $b$. To this end it would be
very interesting to extend this type of analysis to a wider set of pure gauge theories.
For instance, one can consider the Yang-Mills theory with gauge group $Sp(2)$~\cite{Hol03,Hol03b}
which has the same center $\Z (2)$ as the group $SU(2)$ studied in this paper. 

\vskip1.0cm
{\bf  Acknowledgments.}
We acknowledge useful discussions with F.~Gliozzi, J.~Juge, J.~Kuti, C.~Morningstar,
U.-J.~Wiese. This work was partially supported by the European Commission TMR programme
HPRN-CT-2002-00325 (EUCLID) as well as by the Schweizerischer Nationalfond.

\end{document}